\documentclass{jfm}
\usepackage{graphicx,color}
\usepackage{epstopdf, epsfig}
\usepackage[usenames,dvipsnames]{xcolor}    

\renewcommand{\d} {\mathrm{d}}   

\newcommand{\vct}[1] {\ensuremath{\boldsymbol{#1}}} 

\newcommand{\vb} {\vct{b}}
\newcommand{\vj} {\vct{j}}
\newcommand{\vv} {\vct{v}}
\newcommand{\vx} {\vct{x}}

\newcommand{\vz} {\vct{z}}
\newcommand{\vr} {\vct{r}}
\newcommand{\vk} {\vct{k}}
\newcommand{\vdot}{\vct{\cdot}}

\newcommand{\vB} {\vct{B}}

\shorttitle{Decay of anisotropic MHD}
\shortauthor{Bandyopadhyay, Matthaeus, Oughton \& Wan}

\title{Evolution of similarity lengths in anisotropic
        magnetohydrodynamic turbulence}

\author{Riddhi Bandyopadhyay,\aff{1,2}
  \corresp{\email{riddhib@udel.edu}}
  William H.\ Matthaeus,\aff{1,2}
  Sean Oughton,\aff{3}
 \and Minping Wan\aff{4}
 }

\affiliation{\aff{1} Department of Physics and Astronomy, University
        of Delaware, Newark, DE 19716, USA
\aff{2} Bartol Research Institute, University of Delaware, Newark, DE
        19716, USA
\aff{3} Department of Mathematics and Statistics, University of
        Waikato, Hamilton 3240, NZ
\aff{4} Department of Mechanics and Aerospace Engineering, Southern
        University of Science and Technology,
        Shenzhen, Guangdong 518055, People's Republic of China}

\begin{document}

\maketitle

\begin{abstract}
In an earlier paper \citep{Wan2012JFM}, the authors showed that a
similarity solution for anisotropic incompressible 3D
magnetohydrodynamic (MHD) turbulence,
in the presence of a uniform mean magnetic field  $\vB_0$,
exists if the ratio of parallel to perpendicular
   (with respect to $\vB_0$) similarity length
scales remains constant in time.   This conjecture appears to be a
rather stringent constraint on the dynamics of decay of the
energy-containing eddies in MHD turbulence.   However, we
show here, using direct numerical simulations,
that this hypothesis is indeed satisfied in incompressible MHD
turbulence.
After an initial transient period, the ratio of
parallel to perpendicular length scales fluctuates around a steady
value during the decay of the eddies.   We show further that a
Taylor--K\'arm\'an-like similarity decay holds for MHD turbulence in
the presence of a mean magnetic field.
{The effect of different parameters, including Reynolds number,
  DC field strength, and cross-helicity, on the nature of similarity decay
  is discussed.}
\end{abstract}

\begin{keywords}
Authors should not enter keywords on the manuscript, as these must be
chosen by the author during the online submission process and will
then be added during the typesetting process (see
http://journals.cambridge.org/data/\linebreak[3]relatedlink/jfm-\linebreak[3]keywords.pdf
for the full list)
\end{keywords}

        \section{Introduction}
        \label{sec:intro}

Turbulence is a ubiquitous phenomenon that spans a broad range of
temporal and spatial scales.   In a turbulent system energy is
transferred from large
   {(or ``energy-containing'')}
eddies to small eddies,
ultimately resulting in production of internal energy or heat
by dissipation.   This process of energy cascade is observed in
turbulent neutral fluids as well as turbulent plasmas.   The rate of
energy decay in a turbulent system is
        {both}
an interesting problem
        \citep{Kolmogorov1941c} and
  {also an important practical one.}
In laminar flows the rate of energy loss is
determined by the molecular viscosity of the fluid, but in a turbulent
system the energy dissipation rate appears to become independent of
viscosity and
approach a non-zero value as the fluid becomes increasingly
turbulent \citep{Onsagar1949d}.
  \cite{Taylor1938PRSLA} gave an
empirical expression for        {the}  energy decay  {rate}
of turbulent neutral fluids.
{This analytical expression can be obtained from
  the work of von K\'arm\'an \& Howarth (\citeyear{Karman1938PRSL}) by
  assuming the preservation of the shape of the
  two-point correlation function during turbulent decay
   \citep{Dryden1943QAM}}.

Energy decay in plasmas is a more complicated
problem.   Magnetohydrodynamic (MHD) theory is the simplest extension
of hydrodynamic turbulence theories to conducting fluids
  {and the}
     decay of energy-containing eddies in MHD turbulence
  {has been the subject of various studies}
      \citep[e.g.,][]{Hossain1995PoF,Hossain1996AIP,
                        Linkmann2015PRL,Linkmann2017PRE}.
  {Note that the}
presence of
an external mean magnetic field,  $ \vB_0 $,
makes the problem of energy decay in
MHD more complex because the mean field introduces anisotropy in the
system
        \citep{Robinson1971PoF,Montgomery1981PoF,Montgomery1982PS,
                Shebalin1983JPP,Oughton1994JFM,Hossain1995PoF}.

Energy cascade through the MHD inertial scales is given by a
Kolmogorov--Yaglom \citep{Kolmogorov1941c,MoninYaglom-vol2}
third-order law extended to MHD
        \citep{Politano1998GRL,Politano1998PRE}.
On average,
the energy transfer rate associated with cascading of
excitation from the large scales
  (determined by the von K\'arm\'an-type phenomenology)
and the energy cascade rate associated with inertial range scales
  (given by the third-order law)
are expected to be in agreement
        \citep{Kolmogorov1941c,Batchelor1953book,Bandyopadhyay2018bApJ}.

The problem of
similarity decay in isotropic and anisotropic incompressible MHD
turbulence is studied in detail by \cite{Wan2012JFM}, where it is
shown that similarity solutions are possible in MHD. {However, unlike the situation for isotropic neutral fluids, universality, even at asymptotically high Reynolds number, is not expected for MHD due to potential variation of other parameters such as	magnetic Prandtl number, cross helicity, magnetic helicity, and	Alfv\'en ratio. Thus one might anticipate that in MHD the conditions for obtaining similarity solutions are more restrictive. Indeed, \cite{Wan2012JFM} shows, analytically, that a similarity
solution for MHD fluid with a mean magnetic field is possible only if,
during the similarity decay, the similarity length scale parallel to
the mean field remains proportional to the similarity length scale
perpendicular to the mean field.}
As far as we are aware,
whether the two length scales remain in constant proportion
has not yet been tested in simulations or experiments.
This motivates the present study.   We will discuss results from a set
of (spectral method) numerical experiments of MHD turbulence
that examine the dynamical behaviour of the
ratio of the parallel and perpendicular correlation scales.   This
enables us to assess whether or not a similarity decay phase
occurs in anisotropic 3D MHD in the presence of a DC magnetic field
        \citep{Wan2012JFM}.
The results confirm, perhaps surprisingly, that the required
condition for similarity decay of anisotropic MHD can be satisfied.

        \section{Theory} \label{sec:theory}

In this section we briefly review the
        \cite{Wan2012JFM}
derivation of similarity decay
phenomenology for anisotropic MHD with a mean magnetic field.
As in that work, we take the mass density to be constant and set it
to unity.   We denote the fluctuating velocity and magnetic fields by
        $\vv$
and     $\vb$, respectively.
Without loss of
generality, we take the mean magnetic field as
        $ {\vB}_{0} = B_{0}\hat{\vz}$.
All magnetic fields are converted to Alfv\'en
units.   The equations of incompressible MHD can be written in terms of
Elsasser variables for the fluctuations,
        $ {\vz}^{\pm} = \vv\pm\vb $,
as
\begin{eqnarray}
\frac{\partial {\vz}^{\pm}}{\partial t} = - \vz^{\mp}
  \vdot \nabla {\vz}^{\pm} \pm {\vB}_{0} \vdot \nabla
  {\vz}^{\pm} - \nabla P +	 \nu \nabla^2
  {\vz}^{\pm},\label{eq:mhd_zpm}
\end{eqnarray}
where $P$ is the total (magnetic+kinetic) pressure,  and $\nu$ is the
viscosity,      {for simplicity}
assumed to be equal to the resistivity here{in}.

The second-order correlation tensors for the corresponding  Elsasser
fields are defined as
\begin{eqnarray}
   R^{\pm}_{ij}(\vr,t)
        =
   \langle z_{i}^{\pm} (\vx,t) z_{j}^{\pm}(\vx+\vr,t) \rangle,
 \label{eq:Rij}
\end{eqnarray}
where $\langle \cdots \rangle$ denotes an ensemble average.
  {Using the MHD equations}
one can derive the following
equation for
        {the}
time evolution of the   {traced}
second-order correlation function,
\begin{eqnarray}
   \frac{\partial}{\partial t}
    R^{\pm}_{ii}(\vr,t)
        =
   -\frac{\partial}{\partial r_{k}}
        \left[ \hat{Q}^{\pm}_{k}(\vr,t) -
                \hat{Q}^{\pm}_{k}(-\vr,t) \right]
     +  2 \nu \frac{\partial^{2} R^{\pm}_{ii}}
                   {\partial r_{k} \partial r_{k}}
 \label{eq:vkh3}.
\end{eqnarray}
Here
\begin{eqnarray}
  \hat{Q}^{\pm}_{k}(\vr,t)
        =
    \langle   {z^{\mp}_{k}(\vx+\vr,t)}
               z^{\pm}_{i}(\vx,t)
               {z^{\pm}_{i}(\vx+\vr,t)}
    \rangle
 \label{eq:3corr}
\end{eqnarray}
is a triple correlation.
{Interestingly, the mean magnetic field ${\vB}_0$
  does not appear \emph{explicitly} in (\ref{eq:vkh3}), despite the
  well-known fact that these correlation functions (and their Fourier
  transforms, the Elsasser energy spectra) do display anisotropy
  relative to $ {\vB}_0 $.
  In fact, the first explicit appearance of a DC
  magnetic field in the correlation function hierarchy is in the
  equation for evolution of the \emph{third}-order correlations
        \citep{Wan2012JFM,OughtonEA13}.
  Consequently, the dynamical influence of
  ${\vB}_0$ is exerted on (\ref{eq:vkh3}) through the
  structure of the third-order correlations.   Therefore, the limit of
  large ${\vB}_0$ that permits reduction to quasi-2D MHD must
  occur in those higher order equations, and not directly in
  (\ref{eq:vkh3}).}

{A set of similarity solutions can be derived from these
  equations using a
  heuristic scaling argument,   as shown by von K\'arm\'an \& Howarth
  (\citeyear{Karman1938PRSL}) for hydrodynamic turbulence and
  generalized to MHD by \cite{Wan2012JFM}.   Here, we outline the steps
  only for the case of anisotropic MHD decay in the presence of a mean
  magnetic field.   It is well established that a mean magnetic field
  affects the dynamics of the dissipation rate in a turbulent system
        \citep[e.g.,][]
          {Shebalin1983JPP,Oughton1994JFM,
           Bigot2008PRE,Bigot2008PRL}.}
Without loss of generality, we
allow for non-zero cross helicity
        $ H_c = \langle \vv\vdot\vb \rangle$.
The zero cross-helicity case is recovered as a special solution.

Assuming the system to be axisymmetric with
respect to the mean field direction $ \hat{\vz} $,
we write the second-order correlation functions in the form
\begin{eqnarray}
  R^{\pm}_{ii}(\vr,t) = R^{\pm}_{ii}(r_{\parallel},r_{\perp},t)
\label{eq:Raniso},
\end{eqnarray}
where   $ r_{\parallel} = \vr\vdot\hat{\vz} $
and
        $ r_{\perp} = |\vr-r_{\parallel}\hat{\vz}| $.
Clearly,
        $ r_{\parallel}$
and     $ r_{\perp}$
are equivalent to the height $(z)$ and radial $(s)$
coordinates in the usual cylindrical polar coordinate system
        $(s,\phi,z) $.
{Using the theory of axisymmetric tensors
        \citep{Batchelor1953book,Politano2003PRE},
the triple correlations
can be written as}
\begin{eqnarray}
   \hat{Q}^{\pm}_{k}(\vr,t)
        \;=&
     A^{\pm} (r_{\parallel},r_{\perp},t) \hat{r}_{k}
         & +\;
    C^{\pm} (r_{\parallel},r_{\perp},t) \hat{\vz}
 \label{eq:Qpdecomp},\\
   \hat{Q}^{\pm}_{k}(-\vr,t)
         \;=& -A^{\pm} (-r_{\parallel},r_{\perp},t) \hat{r}_{k}
          &+\; C^{\pm} (-r_{\parallel},r_{\perp},t) \hat{\vz}
 \label{eq:Qmdecomp}.
\end{eqnarray}
Inserting expressions (\ref{eq:Qpdecomp})--(\ref{eq:Qmdecomp}) into
        (\ref{eq:vkh3})
yields
\begin{eqnarray}
  \partial_{t} R^{\pm}_{ii}
        =
      -\bigg(\frac{\partial A^{\pm}_{2}}{\partial r_{\perp} } \frac{
  r_{\perp} }{r} + \frac{\partial A^{\pm}_{2}}{\partial r_{\parallel}
  } \frac{ r_{\parallel} }{r} + \frac{2 A^{\pm}_{2} }{r} +
  \frac{\partial C^{\pm}_{2} }{\partial r_{\parallel} } \bigg) + 2 \nu
  \bigg( \frac{\partial^{2} R^{\pm}_{ii} }{\partial r^{2}_{\perp} } +
  \frac{1}{r_{\perp} } \frac{\partial R^{+}_{ii} }{\partial r_{\perp}
  } + \frac{\partial^{2} R^{\pm}_{ii} }{\partial r^{2}_{\parallel} }
  \bigg)
 \label{eq:vkh_aniso},
\end{eqnarray}
where
\begin{eqnarray}
   A^{\pm}_{2} (r_{\parallel},r_{\perp},t )
  &=&
   A^{\pm} (r_{\parallel},r_{\perp},t) + A^{\pm}
      (-r_{\parallel},r_{\perp},t),
 \label{eq:A2}
\\
   C^{\pm}_{2} (r_{\parallel},r_{\perp},t )
  &=&
   C^{\pm} (r_{\parallel},r_{\perp},t) - C^{\pm}(-r_{\parallel},r_{\perp},t),
 \label{eq:C2}
\end{eqnarray}
{with}
  $ A^{\pm}_{2} (-r_{\parallel},r_{\perp},t )
        =
    A^{\pm}_{2} (r_{\parallel},r_{\perp},t )$
and
  $ C^{\pm}_{2} (-r_{\parallel},r_{\perp},t )
        =
  C^{\pm}_{2} (r_{\parallel},r_{\perp},t ) $.

{Following
        von K\'arm\'an \& Howarth (\citeyear{Karman1938PRSL})
 and
        \cite{Wan2012JFM}
 we assume}
\begin{eqnarray}
  R^{+}_{ii}(\vr,t)
        &=&
    Z^{2}_{+} f(\eta_{\parallel},\eta_{\perp}) ,
 \label{eq:Rsim}
\\
  A^{+}_{2}(\vr,t)
        &=&
    Z_{-} Z^{2}_{+} a(\eta_{\parallel},\eta_{\perp}) ,
 \label{eq:Asim}
\\
  C^{+}_{2}(\vr,t)
        &=&
    Z_{-} Z^{2}_{+} c(\eta_{\parallel},\eta_{\perp}) .
 \label{eq:Csim}
\end{eqnarray}
{introducing the normalized variables}
        $ \eta_{\parallel}=r_{\parallel}/L^{+}_{\parallel}(t)$
and
        $ \eta_{\perp}=r_{\perp}/L^{+}_{\perp}(t) $,
and {the shorthand notation}
        $ Z^{2}_{\pm} =
        R^{\pm}_{ii}(0,t) = \langle |\vz_{\pm}|^2 \rangle $.

{Using
        (\ref{eq:Rsim})--(\ref{eq:Csim}), in (\ref{eq:vkh_aniso})
  we obtain}
\begin{eqnarray}
   \bigg\{ \frac{\mathrm{d}Z^{2}_{+} }{\mathrm{d}t}\bigg\} [f]
 - \bigg\{ \frac{Z^{2}_{+} }{L^{+}_{\parallel} }
           \frac{\mathrm{d}L^{+}_{\parallel}}{\mathrm{d}t} \bigg\}
    \bigg[ \frac{\partial f}{\partial \eta_{\parallel}}
            \eta_{\parallel}
    \bigg]
  - \bigg\{ \frac{Z^{2}_{+}}{L^{+}_{\perp}}
            \frac{\mathrm{d}L^{+}_{\perp}}{\mathrm{d}t} \bigg\}
    \bigg[ \frac{\partial f}{\partial \eta_{\perp}}
            \eta_{\perp}
    \bigg]
 &\nonumber
\\ + \bigg\{ \frac{Z_{-} Z^{2}_{+} }{L^{+}_{\perp} } \bigg\}
     \bigg[   \frac{1}{\sqrt{\eta^{2}_{\perp}
            + \gamma^{2} \eta^{2}_{\parallel} }}
       \bigg( \frac{\partial a}{\partial \eta_{\perp} } \eta_{\perp}
              + \frac{\partial a}{\partial \eta_{\parallel} } \eta_{\parallel}
              + 2a
       \bigg)
     \bigg]
 +&\nonumber
     \bigg\{ \frac{Z_{-} Z^{2}_{+} }{L^{+}_{\parallel} } \bigg\}
     \bigg[ \frac{\partial c}{\partial \eta_{\parallel} } \bigg]
\\
    -\bigg\{2\nu \frac{Z^{2}_{+}}{{L^{+}_{\parallel}}^{2} } \bigg\}
      \bigg[ \frac{\partial^{2} f}{\partial {\eta_{\parallel}}^{2} }
      \bigg]
     - \bigg\{ 2\nu \frac{Z^{2}_{+}}{{L^{+}_{\perp}}^{2} } \bigg\}
       \bigg[ \frac{\partial f}{\partial \eta_{\perp} }
              \frac{1}{\eta_{\perp} }
              \frac{\partial^{2} f}{\partial {\eta_{\perp}}^{2} }
       \bigg] = 0
 \label{eq:vkh_big},&
\end{eqnarray}
where
        $ \gamma = L^{+}_{\parallel}/L^{+}_{\perp} $.
We assume here that
the ``$+$'' and ``$-$'' variables are independent of each other.
For ease of identification
we have written all the terms that are explicitly dependent on time
inside curly brackets:        $ \{ \cdots \} $.
Terms that do not explicitly depend on time are written inside
square brackets:  $ [ \cdots ] $.
  There are two points to note here.
   First, because
        $\eta_{\parallel} $     and     $ \eta_{\perp} $ are functions
   of time, the square-bracketed terms will in general have
   \emph{implicit} time dependence.
   Second, a priori one would expect
   the variable
        $ \gamma = L^{+}_{\parallel} / L^{+}_{\perp} $
   to be time dependent.   Thus, the claim that the square-bracketed
   terms lack explicit time dependence will only be true if $\gamma $
   is constant.
The dynamical relevance of this
constraint is the primary focus of this study.

As an aside,
we note that this requirement for similarity solutions may
not pertain to the asymptotic case of
        $ \gamma^{2} \eta^{2}_{\parallel} \gg \eta^{2}_{\perp}$,
  i.e., $ r^{2}_{\parallel} \gg r^{2}_{\perp}$.
In
this circumstance, the fourth term of (\ref{eq:vkh_big}) can be
separated into a time-dependent part and a time-independent part,
regardless of the behaviour of $\gamma$, in the asymptotic
limit.   Physically, this limit would be relevant to phenomena or
structures that are strongly elongated along the parallel direction.   A
similarity solution might then exist  without the need for $ \gamma = $constant.

Without pursuing the above mentioned limits here, and assuming that
$\gamma$ remains constant in time, we can gather all
the terms inside curly brackets in (\ref{eq:vkh_big})
and set them proportional to each other.
Proceeding accordingly, we can write
\begin{eqnarray}
   \frac{\mathrm{d}Z^{2}_{+} }{\mathrm{d}t}
        \propto
   \frac{Z^{2}_{+}}{L^{+}_{\perp}}
   \frac{\mathrm{d}L^{+}_{\perp}}{\mathrm{d}t}
        \propto
   \frac{Z_{-} Z^{2}_{+} }{L^{+}_{\perp} }      \label{eq:prop},
\end{eqnarray}
so that
\begin{eqnarray}
  \frac{\mathrm{d}L^{+}_{\perp}}{\mathrm{d}t}
        &=&
    \beta^{+} Z_{-}\label{eq:vkdec1},    \\
  \frac{\mathrm{d}Z^{2}_{+} }{\mathrm{d}t}
        &=&
   - \alpha^{+} \frac{Z_{-} Z^{2}_{+}}{L^{+}_{\perp}}   \label{eq:vkdec2},
\end{eqnarray}
where   $ \beta^{+}$ and $\alpha^{+}$
are both positive time-independent constants.
 {In deriving
         (\ref{eq:vkdec1})--(\ref{eq:vkdec2})
   we have ignored the terms containing the viscosity $\nu$ in
         (\ref{eq:vkh_big}),
  due to the assumption $\nu \ll 1$; i.e., high Reynolds number.}
The
        ``$-$''
versions of (\ref{eq:vkh_big})--(\ref{eq:vkdec2})
are analogous.

{Equations (\ref{eq:vkdec1}) and (\ref{eq:vkdec2}) can be
  heuristically derived from dimensional analysis and modelling
    \citep[e.g.,][]{Dobrowolny1980PRL, Hossain1995PoF, Biskamp2001PoP}.
  The
  derivation presented here and in \cite{Wan2012JFM} highlights the
  underlying assumptions and limitations of these solutions.   For example,
  the derivation relies on the assumption of similarity, i.e., that the
  two-point correlation function maintains its shape during the
  decay.   Moreover, the requirement that $\gamma$ needs to remain
  constant in time is manifested through this analysis.}

Equations (\ref{eq:vkdec1}) and (\ref{eq:vkdec2}) are exactly satisfied
if the solutions obey the conservation law
\begin{eqnarray}
   Z_{+}^{(2 \beta^+ / \alpha^+) }  L^{+}_{\perp}
        =
    \mathrm{constant}.
 \label{eq:cons-law}
\end{eqnarray}
For the long time
behaviour of $Z_{+}$ and $L^{+}_{\perp}$, one expects,
on the basis of physical arguments for decaying turbulence
        \citep{Matthaeus1996JPP},
that
\begin{eqnarray}
	\alpha^{+} \ge \beta^{+} \label{eq:ratio}.
\end{eqnarray}
We now test these hypotheses using spectral simulations.

        \section{Simulations} \label{sec:sim}

To test the hypothesis that
  {the Elsasser energies and correlation lengths of (unforced)}
MHD turbulence evolve according to
von K\'arm\'an--Howarth
similarity decay {laws}---equations~(\ref{eq:vkdec1})--(\ref{eq:vkdec2})
   and their `minus' partners---and
 {which also requires}
that the ratio of
the parallel and perpendicular characteristic lengths does not change
in time, we carry out a set of   {incompressible}   MHD simulations
with a mean magnetic field,
        $ B_0 \hat{\vz} $.

All runs are initialized with kinetic and magnetic spectra
proportional to $ 1/[1+(k/k_{0})^{11/3}]$, with $k_{0}=4$ and  only
the Fourier modes within the band $1 \le k \le 15$ excited.
The initial total energy is always normalized to one.
Correlation
lengths are small compared to the total box length for all runs.
Table~\ref{tab:sim}
contains a summary of the simulation parameters used for this study.
  {Although we are here mainly concerned with
    anisotropy induced by a DC magnetic field, for context we also
    include results from an isotropic simulation that lacks a global
    DC field (run11).}

\begin{table}
	\begin{center}
		\def~{\hphantom{0}}
		\begin{tabular}{lcccccc}
			Simulation  & $N^{3}$ & $B_0$   &   $\nu$ & $\sigma_{c}$ & $r_{\mathrm{A}}$ & $\d t$ \\[3pt]
			run0   & $256^{3}$ & 0.5 & 0.002 & 0.0 & 1.0 & 0.001 \\
			run1   & $256^{3}$ & 1 & 0.002 & 0.0 & 1.0 & 0.001 \\
			run2   & $256^{3}$ & 1 & 0.002 & 0.5 & 1.0 & 0.001 \\
			run3   & $256^{3}$ & 1 & 0.002 & 0.0 & 0.5 & 0.001 \\
			run4   & $256^{3}$ & 2 & 0.002 & 0.0 & 1.0 & 0.0005 \\
			run5   & $256^{3}$ & 3 & 0.002 & 0.0 & 1.0 & 0.0004 \\
			run6   & $256^{3}$ & 2 & 0.002 & -0.5 & 1.0 & 0.0005 \\
			run7   & $256^{3}$ & 4 & 0.002 & 0.0 & 1.0 & 0.00025 \\
			run8   & $256^{3}$ & 2 & 0.002 & 0.0 & 2.0 & 0.0005 \\
			run9   & $256^{3}$ & 2 & 0.002 & 0.8 & 2.0 & 0.0005 \\
			run10   & $512^{3}$ & 1 & 0.0005 & 0.0 & 1.0 & 0.0005 \\
			run11   & $256^{3}$ & 0 & 0.002 & 0.5 & 1.0 & 0.0005 \\
		\end{tabular}
		\caption{Simulation parameters for spectral
			simulations: grid size $N^{3}$, the mean magnetic
			field strength $B_{0}$, viscosity $\nu$, initial
			normalized cross-helicity $\sigma_c = 2H_c/E$, initial
			Alfv\'en ratio $r_{\mathrm{A}} = E_v / E_b$,
                        timestep $\d t$.}
		\label{tab:sim}
	\end{center}
\end{table}

We numerically solve (\ref{eq:mhd_zpm}) in a periodic box using a
pseudo-spectral solver without any external forcing.   All the variables
are expanded in a Fourier basis with transfer between real space and
Fourier space performed using a Fast Fourier Transform (FFT).   We
use the second-order Runge--Kutta (RK2) scheme for time-stepping, and
the  $2/3$ rule for dealiasing.
To ensure accuracy of the dissipation rates and spectra we require
that $k_{\mathrm{max}} \zeta > 1 $ for all simulations
        \citep{Wan2010bPoP,Donzis2008PoF}.
Here $k_{\mathrm{max}}$ is the maximum
resolved wavenumber and $\zeta$ is the Kolmogorov {dissipation} length scale.

  {For strong mean field, the simulations can be performed
    in non-cubic boxes, provided the parallel cascade (in addition to
    the perpendicular cascade) is well resolved
    \citep{Oughton2004PoP}.   For a well-resolved case, a non-cubic
    simulation domain is not expected to modify the dynamics in
    incompressible MHD \citep{Bigot2008PRL}.
    We employ a cubic periodic box for all runs discussed herein.}

        \section{Results} \label{sec:result}

To study the  {decay}  dynamics of the energy-containing eddies
we compute, at each timestep,
the ``Elsasser energies''   $ Z_{+}^2 $  and $ Z_{-}^2 $
and their characteristic lengths along each Cartesian coordinate
direction.
  {The latter are calculated from the two-point
    correlation functions
  (see
(\ref{eq:Rij})) as
\begin{eqnarray}
    L_{x}^{\pm}
        =
   \frac{1}{Z^{2}_{\pm}} \int_{0}^{\infty} R^{\pm} (r,0,0) \, \mathrm{d}r ,
 \label{eq:Lx}
\end{eqnarray}
and similarly for the $y$ and $z$ components.} Here, $R^{\pm}=R^{\pm}_{ii}$
are the trace of the correlation tensors.
{In Fourier space,
we can equivalently define the length scales as
\begin{eqnarray}
    L_{x}^{\pm}
        =
   \frac{\pi}{Z^{2}_{\pm}}
        \sum_{k_{y}, k_{z}} |{\vz}^{\pm}(k_{x}=0,k_{y},k_{z})|^{2} .
 \label{eq:Lx_spec}
\end{eqnarray}
So, the length scales $(L_{x}^{\pm})$ are proportional to the
reduced spectrum evaluated at zero wavenumber: $ E^{\mathrm{red}}_{x}(k_x=0) $.}
  {We define
\begin{eqnarray}
  L_{\parallel}^{\pm}
        &=&
  \L_{z}^{\pm}          \label{eq:Lpar},\\
  L_{\perp}^{\pm}
        &=&
  \sqrt{  \frac{{(L_{x}^{\pm})}^2  +  {(L_{y}^{\pm})}^2} {2} }
\label{eq:Lperp}.
\end{eqnarray}}
The factor of $1/2$ in the definition of $L_{\perp}^{\pm}$ is used
because there are two independent components in the perpendicular plane
        \citep[e.g.,][]{Oughton2005NpG}.
With this definition we usefully have
        $ L_{\perp} \approx L_{\parallel} $
for the isotropic case,
when
        $L_{x} \approx L_{y} \approx L_{z}$.

\begin{figure}
	\begin{center}
		\includegraphics[scale=1.0]{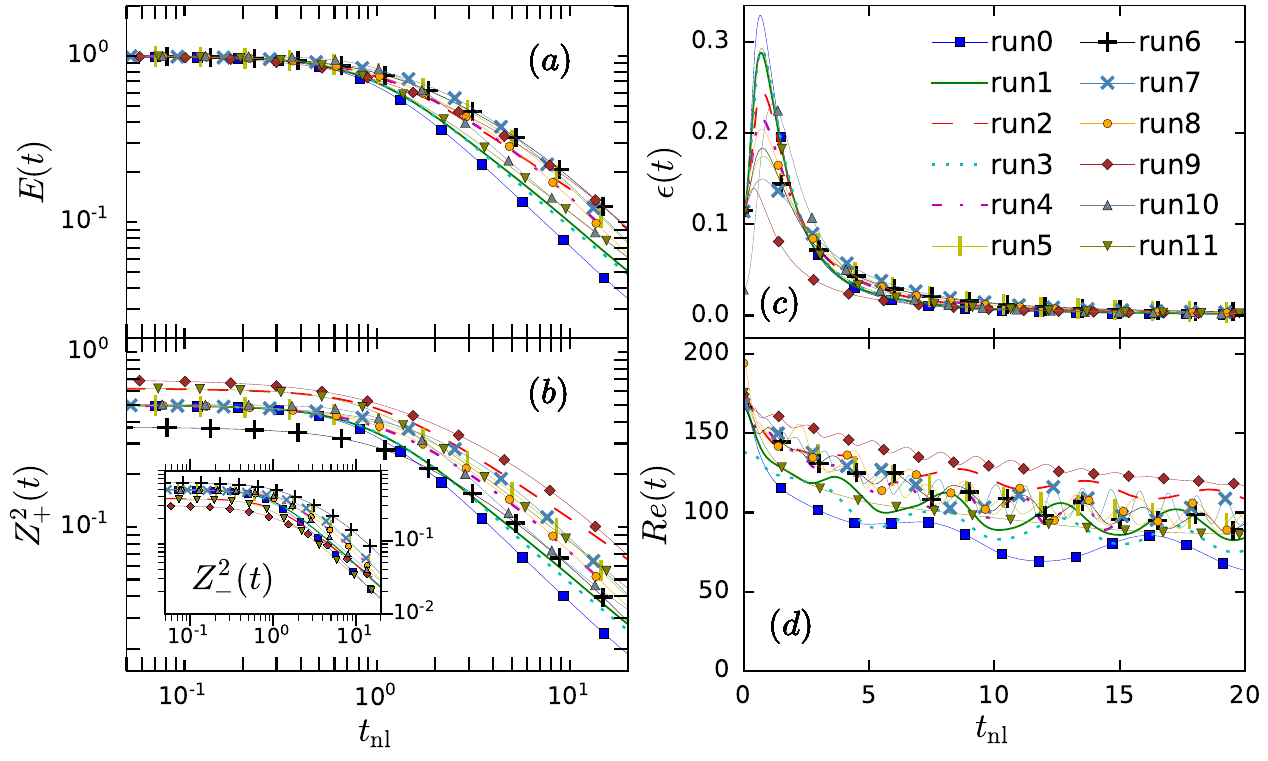}
		\caption{Time history of
                 (a) {the total fluctuation energy $E$,}
                 (b) {the Elsasser energies $Z_{\pm}^2$ with ``$-$'' variables shown in the inset,}
                 (c) the mean energy dissipation rate $\epsilon = \nu
                  \langle j^2 + \omega^2 \rangle$, and
                 (d) fluid Reynolds number $\Rey$ for the runs listed in
                        table~\ref{tab:sim}.
                  All quantities are defined in the text.}
		\label{fig:energy}
	\end{center}
\end{figure}

  {Figure~\ref{fig:energy} shows the time histories of the
    total fluctuation energy $E$ (magnetic$+$kinetic),
    Elsasser energies $Z_{\pm}^2$,
    mean energy dissipation rate
        $ \epsilon = \nu  \langle j^2 + \omega^2 \rangle $,
    and fluid Reynolds number $\Rey$ for
    all the runs in table~\ref{tab:sim}.   Note that these quantities
    are associated with the fluctuations and, in particular, the calculation
    of Elsasser variables, their energies, and the total fluctuation
    energy does not include the contribution from the DC magnetic
    field, ${\vB}_0$.}
Here, $\vct{\omega}=\mathbf{\nabla} \times \vv$ is the
vorticity and $\vj=\mathbf{\nabla} \times \vb$ is the
current density.   The time axis is plotted in units of the initial
nonlinear time
or ``eddy turnover'' time,
        $ t_{\mathrm{nl}} = t/\tau_{\mathrm{nl}} $.
  {We employ the following definition for the initial eddy
    turnover time
\begin{eqnarray}
  \tau_{\mathrm{nl}}
        =
  \frac{ [L^{+}(0) + L^{-}(0)]/2}
       { \sqrt{{Z_{+}^2(0)}  +  {Z_{-}^2(0)}}} ,
 \label{eq:tnl}
\end{eqnarray}}
where $L^{\pm}(0)$ are the initial correlation lengths, and
{
\begin{eqnarray}
	L^{\pm} (t)
	=
	\frac{\pi}{Z^{2}_{\pm}(t)}
          \sum_{\vk} \frac{ |{\vz}^{\pm}(\vk,t)|^2 }
	                       { |\vk| } .
	\label{eq:Lpm}
\end{eqnarray}}
{These are straightforward MHD generalisations of the
  classical definition
  of the correlation length or integral scale
        \citep{Batchelor1953book, Linkmann2015PRL, Bandyopadhyay2018PRX}
\begin{eqnarray}
   L_{\mathrm{int}}
        =
   \frac{\pi}{E^v} \sum_{\vk} \frac{E^v(\vk)}
                                       {|\vk|} .
 \label{eq:Lint}
\end{eqnarray}
Note that we do not recover the directional length scales of
  (\ref{eq:Lx}, \ref{eq:Lx_spec})
by simply replacing $\vk$ with $k_x$ in (\ref{eq:Lpm}).}
The fluid
Reynolds number is defined as
        $ \Rey  =  v^{\prime} L_{\mathrm{int}}/\nu$,
where   $ v^{\prime}$
denotes the     {(average) component}   rms speed
with
        $ E^v = 3(v^{\prime})^2 /2 $.
Here, $E^v(\vk)$ is the modal kinetic energy spectrum and
$E^v$ is the total kinetic energy.

{
Panels (a) and (b) of figure~\ref{fig:energy} indicate
that, for all runs considered,
a powerlaw (in time) is a reasonable approximation to
the decay of both the total fluctuation energy and the
Elsasser energies, after a few nonlinear times. This behaviour is expected for von K\'arm\'an-Howarth similarity
decay \citep{Matthaeus1996JPP}. Not all runs have the identical power-law slope and a full 
explanations for these slight differences is yet to be obtained.  However, it 
is clear that the decay is (approximately) self-similar at these later 
times. During these $ t  \gtrsim  5  \tau_{\mathrm{nl}} $
periods the dissipation rates are much smaller than the
peak values but are also only slowly decreasing;
the Reynolds numbers are also slowly decreasing (with oscillations).
}

\begin{figure}
	\begin{center}
		\includegraphics[scale=1.0]{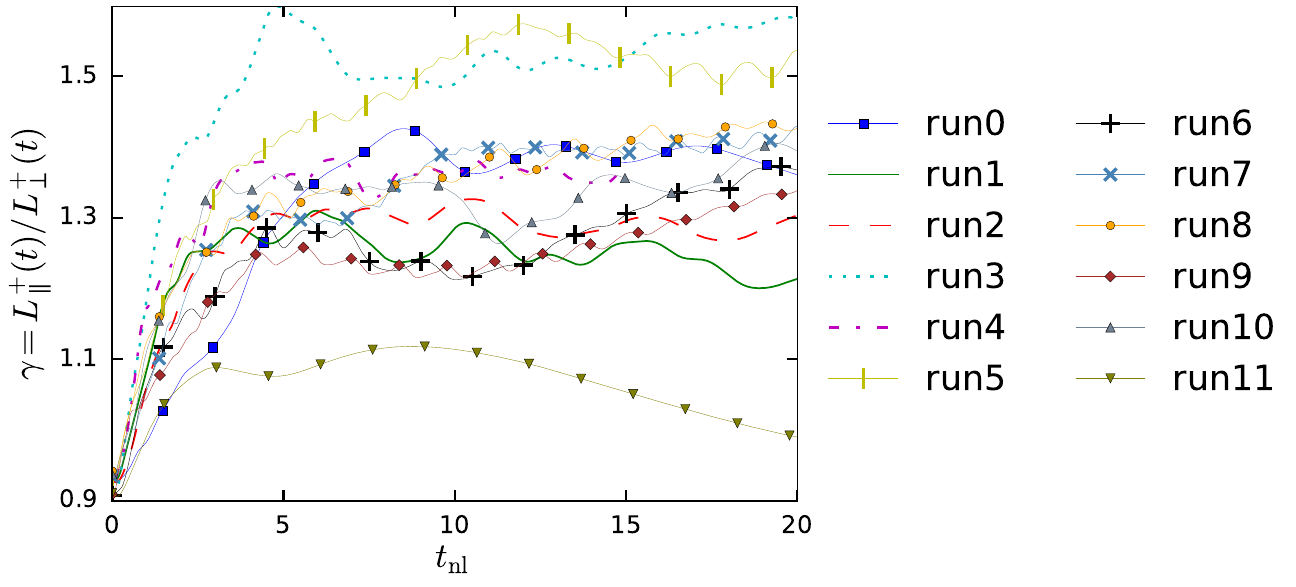}
		\caption{Time evolution of the ratio of
                  parallel to perpendicular correlation lengths,
                        $ \gamma=L^{+}_{\parallel}/L^{+}_{\perp} $,
                  for the runs of table~\ref{tab:sim}.}
		\label{fig:const}
	\end{center}
\end{figure}

Figure~\ref{fig:const} illustrates the time history of the ratio of
the parallel to perpendicular length scales
        $ \gamma = L_{\parallel}^{+}/L_{\perp}^{+}$,
corresponding to the ``$+$'' Elsasser variable.
It is evident that after an initial transient period, the ratio of the
two lengthscales typically saturates to an approximately steady value,
  {with}  fluctuations around     {those values.}

  {A closer inspection of figure~\ref{fig:energy} and
    figure~\ref{fig:const} reveals that the dissipation rate reaches
    its maximum near unit nonlinear time.   However, $\gamma$ values
    saturate at a somewhat later time $t_{\mathrm{nl}}\sim 2-5$.   This
    behaviour can probably be explained by noting that modifying
    the very large lengthscales takes a long time.   The correlation
    lengths may become steady after the lowest wavenumber part of the
    spectrum is well populated.   Since dissipation involves high
    wavenumber regions of the spectrum, where the characteristics
    timescales are much faster than those of the energy-containing
    eddies, it is perhaps not surprising that the dissipation rate
    peaks before $\gamma$ saturates.}

Although $\gamma$ attains different values for different
simulation sets, figure~\ref{fig:const}         {indicates}
that for all cases $\gamma$ remains  {approximately}  stationary for
        {many nonlinear}
times.
Furthermore,
        {for the nonzero mean field cases},
   $L^{+}_{\parallel}$ is always greater than $L^{+}_{\perp}$,
indicating
that the correlation lengths along the mean field are longer than
those perpendicular to it, due to   {the}  cascade preferentially
transferring energy in the perpendicular directions
        \citep{Shebalin1983JPP,Grappin86,
        	Matthaeus1990JGR, Oughton1994JFM, GoldreichSridhar95, Teaca2009PRE}.

Figure~\ref{fig:const} contains the main results of this paper.
 (Plots for
        $\gamma^- = L^-_\parallel / L^-_{\perp} $,
   not shown, are completely analogous.)
Having
established that the ratio of parallel to perpendicular length scales
remain  {(roughly)}       constant in time, we proceed to  examine
whether the proposed von K\'arm\'an similarity decay is satisfied for
MHD fluids in the presence of a global magnetic field.
\begin{figure}
	\begin{center}
		\includegraphics[scale=1.0]{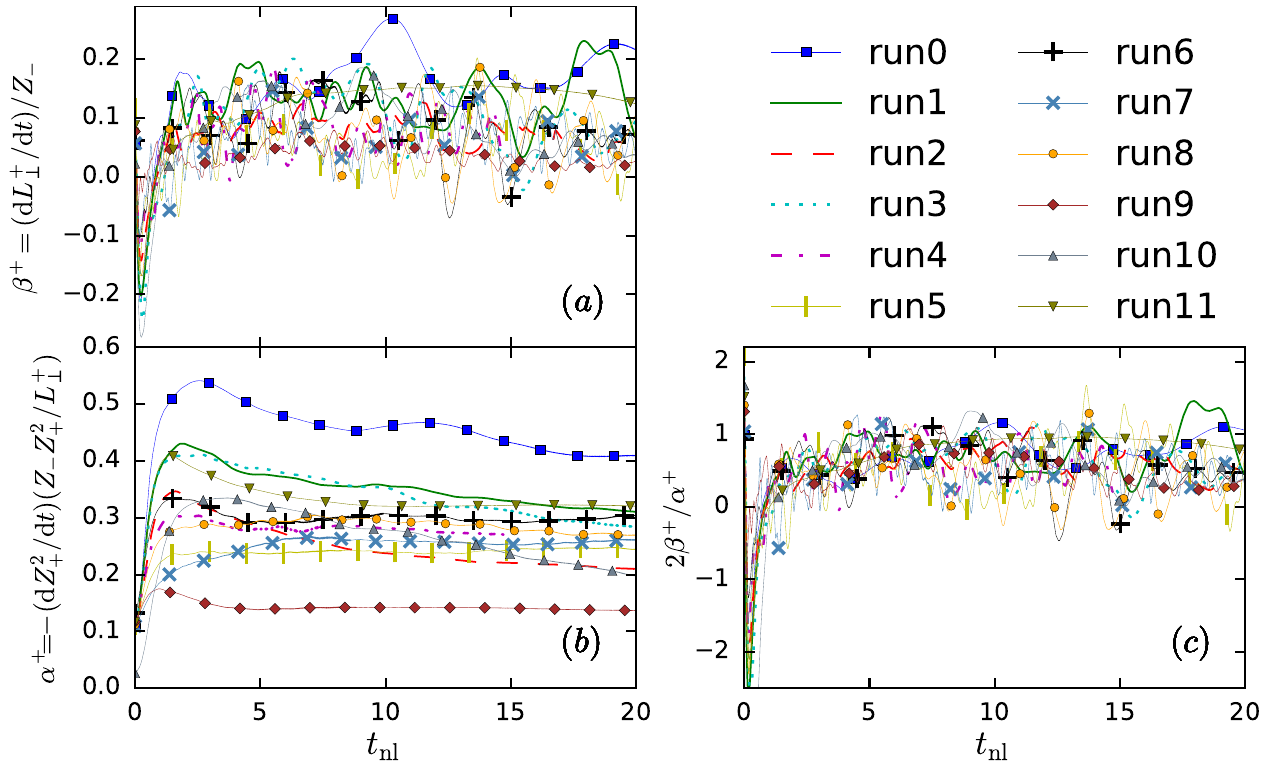}
		\caption{Time evolution of the two von K\'arm\'an
                  constants:
                (a)
                  $ \beta^{+} = (\d L_{\perp}^{+} /\d t)/Z_{-}$.
                (b)
                  $ \alpha^{+} = - (\d Z_{+}^2 /\d t)(Z_{-}
                                Z_{+}^2/L_{\perp}^{+})$, and
                (c) twice the ratio of the two constants,
                        $ 2\beta^{+}/\alpha^{+} $,
                 associated with the family of conservation laws,
                 equation~(\ref{eq:cons-law}).}
		\label{fig:vk}
	\end{center}
\end{figure}
In figure~\ref{fig:vk}, we show the two ``von K\'arm\'an constants''
$\alpha^{+}$ and $\beta^{+}$, corresponding to the ``$+$'' Elsasser
variables, as functions of time.   If the similarity decay
hypothesis is indeed satisfied, these two quantities should maintain
constant values in time.
Figure~\ref{fig:vk}(a)
shows the
rate of change of perpendicular length scale $L_{\perp}^{+}$
normalized to $Z_{-}$, which, if the decay obeys a similarity solution is a
constant, $\beta^{+}$.
Panel (b)
plots the (negative) normalized rate of change of $Z_{+}^2$ as
a function of time; again, if the decay obeys a similarity this will be a
constant, $\alpha^{+}$.
A central
difference scheme is used to evaluate the time derivatives.   It is
clear from the two panels of figure~\ref{fig:vk} that the similarity
decay hypothesis, as proposed in \cite{Wan2012JFM}, is well supported
by the simulation results presented here.

Recall also, from equation~(\ref{eq:cons-law}), that the conserved
quantity associated with the self-similar decay depends on the ratio
        $ 2\beta^{+}/\alpha^{+} $.
  {This ratio is displayed in panel (c) of
        figure~\ref{fig:vk}
     where it can be seen that it attains a steady-state value of
     somewhat less than unity, after an initial adjustment phase.
    From \cite{Dryden1943QAM} and \cite{Karman1949RMP}
   self-similar decay for all scales requires
   $ \alpha^{\pm} = \beta^{\pm} $.
   This situation corresponds to the case
   of decay with constant turbulent viscosity, $Z^{\pm} L^{\pm}=$
   constant, or equivalently decay at constant Reynolds number.
   Clearly, this is not satisfied rigorously in the simulations
   presented here.  Further, the plotted values of
   $ 2\beta^{+} / \alpha^{+}$ appear to eliminate the possibility of
   similarity decay with ${(Z^{\pm})}^2 L^{\pm}=$ constant, physically
   corresponding to the case of constant area under the correlation
   function.}
Although only the
        ``$+$''
Elsasser variables are shown here, the    {results are similar}
for the
        ``$-$''
Elsasser variables.
 {As an aside we note that applications of MHD
   decay phenomenologies within studies of the transport of solar wind
   fluctuations
        \cite[e.g.,][and many subsequent
        papers]{Matthaeus1996JPP,ZankEA96}
   have previously employed both the
        $ \beta / \alpha = 1$ and the $ 2\beta/\alpha = 1 $
   conditions.
 }

{Next, we briefly discuss the effect of anisotropy due to
  the DC magnetic field strength  $ B_0$  and/or the cross-helicity
  strength.  Of particular interest here is the variation of
        $ \gamma = L^+_\parallel / L^+_\perp$
  with $ B_0$ and with the magnitude of
  the initial normalized cross-helicity $\sigma_{\mathrm{c}} $.
  Figure~\ref{fig:g_var} shows the asymptotic values of $\gamma$ for
  all the runs.  These asymptotic values, denoted $\gamma_{*}$,
  are obtained by averaging
  $\gamma$ over the final five nonlinear times for each run.
  Within the limited parameter range scan of
  $ B_0$ and $\sigma_{\mathrm{c}}$
  covered by the
  simulations presented here, it appears that $\gamma_{*}$ initially
  increases with $B_0$ but the effect then saturates for higher values
  of $B_0$.  This behaviour is expected since the mean-field induced
  anisotropy renders the system approximately two-dimensional; i.e.
  $L_{\parallel} > L_{\perp}$.  On the other hand, from panel (b) of
  figure~\ref{fig:g_var}, no clear scaling can be deduced between
  $\gamma_{*}$ and $\sigma_{\mathrm{c}}$.

\begin{figure}
	\begin{center}
		\includegraphics[scale=1]{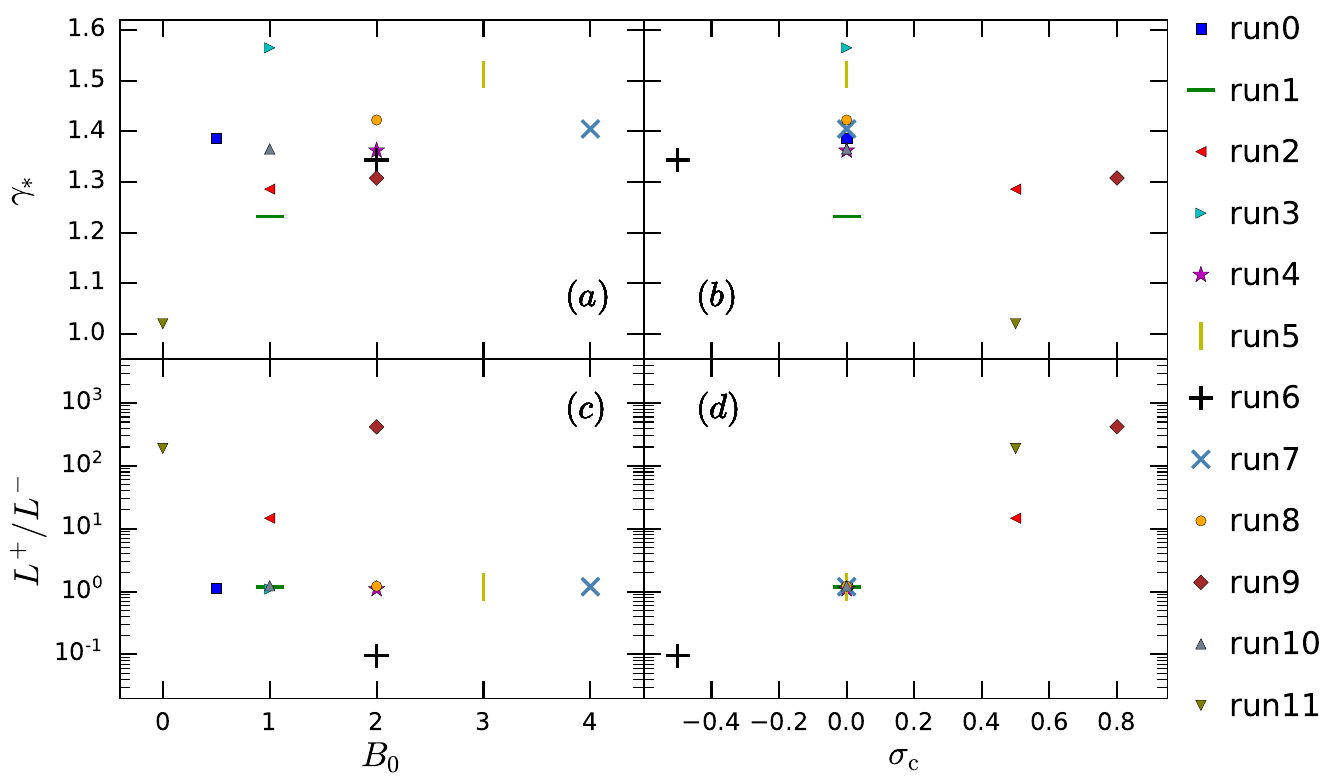}
		\caption{Variation of the asymptotic
                  values of $\gamma = L_{\parallel}^{+}/L_{\perp}^{+}$
                  with (a) mean-magnetic-field strength $B_0$ and
                  (b) the normalized cross-helicity $\sigma_{\mathrm{c}}$,
                  for the runs of table~\ref{tab:sim}.
                  Panels (c) and (d)
                  plot the variation of the ratio $L^{+}/L^{-}$
                  with $B_0$ and $\sigma_{\mathrm{c}}$, respectively.}
		\label{fig:g_var}
	\end{center}
\end{figure}

A related quantity of interest is the ratio of the length scales
corresponding to the $``+"$ and $``-"$ Elsasser variables.  The bottom
two panels of figure~\ref{fig:g_var} illustrate the variation of the
ratio $L^{+}/L^{-}$.  Again, the reported values are the temporal
averages over  the  final five nonlinear time units.  Here, we see that
there is no evident scaling with the mean-field strength $B_0$.
However, panel (d) exhibits  a  rough increasing scaling of
        $ L^{+}/L^{-}$ with $\sigma_{\mathrm{c}}$.
This result is consistent
with the explanation provided by
        \cite{Matthaeus1983PRL} and \cite{Ghosh1988PoF},
who argue that for high cross-helicity
(say, positive), one of the Elsasser fields $({\vz}_{-})$ is weak
and it is almost passively advected towards high wavenumber by the
dominant Elsasser field $({\vz}_{+})$.  This kind of tendency
would result in dissimilar values of the two Elsasser field
length scales $(L^{+} \gg L^{-})$.  From figure~\ref{fig:g_var}, the
zero cross-helicity runs, run 0, 1, 3, 4, 5, 7, 8, 10, maintain
$ L^{+}/L^{-} \approx 1$.
The positive cross-helicity runs, run 2, 9, 11,
show $L^{+}/L^{-} > 1$, with increasing value of $L^{+}/L^{-}$ as
$\sigma_{\mathrm{c}}$ increases.  Run 6 has $\sigma_{\mathrm{c}} < 0$
and consequently $L^{+}/L^{-} < 1$ for this case.}

        \section{Discussion}
        \label{sec:disc}

We have examined the validity of a von K\'arm\'an--Howarth-like
        similarity decay phase
in anisotropic 3D MHD in the presence of an externally
supported DC magnetic field ($ {\vB}_0 $),
as derived in
        \cite{Wan2012JFM}.
An analytic result is that a similarity decay phase is only
followed in an MHD fluid (with a $ {\vB}_0 $)
if the ratio of the parallel to perpendicular characteristic lengthscales, $\gamma=L_{\parallel}^{+} / L_{\perp}^{+}$,
remains  {constant}  in time.
Using numerical simulations, performed with a range of different
parameters, we find that the ratio of
parallel to perpendicular length scales does indeed maintain an
approximately steady value during
the decay of the MHD turbulence, after an initial build-up phase.
This result provides substantial support for the occurrence of
similarity decay of energy in MHD turbulence with a mean field.

Additionally, since the ratio of the parallel and perpendicular
lengthscales maintains a constant value this implies that only one of the
lengthscales evolves independently.
This has useful consequences for global
turbulence-based modelling of the solar wind and other astrophysical
plasmas.    Often in such models a von K\'arm\'an-like phenomenology is
invoked
        \citep[e.g.,][]{Breech2008JGR, OughtonEA11, Usmanov2014ApJ}.
If
the parallel and perpendicular lengths maintain a constant proportion
in the solar wind, it may be sufficient to evolve the lengthscale
along only one direction, simplifying the calculations and possibly
making the computations less expensive.

{The results presented in this paper are important for
  understanding
heating and acceleration of space plasmas such as the solar corona,
solar wind, and magnetospheric plasmas.   The conclusions
{should}
also be useful for understanding and modelling the role of
turbulence in the evolution and dynamics of astrophysical plasmas and
laboratory plasmas.}

{We note that although the runs have the same initial
  conditions, after a few nonlinear times they evolve independently to
  distinct states.  Therefore, we   suggest  that the result, that
  $L_{\parallel} / L_{\perp}$ maintains a steady value, does not
  depend on the large-scale eddies being of the same form in the parallel
  and perpendicular directions.}

{The applicability of the theory for modest Reynolds
  number warrants some discussion.  To arrive at (\ref{eq:vkdec1}) and
  (\ref{eq:vkdec2}), we neglect the two terms proportional to $\nu$
  in (\ref{eq:vkh_big}).  This step can be justified by a simple
  calculation to estimate the order of magnitude of the neglected
  terms compared to the retained terms in (\ref{eq:vkh_big}).  Let
  us compare the two terms
        $ \{\d Z_{+}^2/\d t\}[f]$
  and   $\{2 \nu Z_{+}^2/ {L_{\parallel}^{+}}^2 \}
            [\partial^2 f / \partial \eta_{\parallel}^2 ] $.
  For simplicity, we ignore the notations
  $\parallel$, $\perp$, $\pm$, etc., and assume $f \sim \exp(-\eta) $.
  Then, we can compare the two terms as
\begin{eqnarray}
	\frac{\d Z^{2}}{\d t}: \nu \frac{Z^{2}}{L} \label{eq:comp1}.
\end{eqnarray}
For a consistency check, if we insert the desired solution,
        $ \d Z^2 / \d t  \sim  Z^3 / L$, on the LHS we obtain
\begin{eqnarray}
    \frac{Z^{3}}{L}   &:&   \nu \frac{Z^{2}}{L} .  \label{eq:comp2}
\end{eqnarray}
Noting that $ Z L / \nu \sim \Rey$ this yields
\begin{eqnarray}
		1  : \frac{1}{\Rey}             \label{eq:1oRe}.
\end{eqnarray}
So, from this very crude argument, the theory is expected to hold for
$\Rey \gg 1$.  In practice, one finds that the conditions for a
similarity decay law  are  much less stringent than, say, the conditions
for the Kolmogorv $-5/3$ slope \citep{Karman1949RMP}.
In the
simulations shown here, the lowest value of Reynolds number is around
fifty, $\Rey \sim 50$.  Therefore, in the worst case scenario, the
neglected terms are about $50$ times smaller than the retained terms
in (\ref{eq:vkh_big}).  It is clear from the results that this
level of smallness for the terms proportional to $\nu$
is sufficient to satisfy an approximate similarity decay.}

{However, we can infer the effect of Reynolds number and
  large-scale eddy strength $(\sim Z L)$ by examining results from two
  simulations, run 3 and run 10.  These
  differ only by Reynolds number with run 10 having the larger
     $ \Rey $.
  From figure~\ref{fig:g_var}, run 3 has a higher
  value of $\gamma^{*}$
     (i.e., asymptotic $L_{\parallel}^{+} / L_{\perp}^{+}$ ).
  One factor contributing to the
  different values of $\gamma^{*}$ is probably the different
  grid size in the two simulations.  Further, it is known that
  mean-field-induced anisotropy depends on Reynolds number, so that
  may play a role here.  The ratio $L^{+}/L^{-}$, on the other hand,
  admits almost equal values for the two runs, presumably since the
  cross-helicity is the same for both cases.  The `energy' similarity decay
  constant, $\alpha^{+}$, expectedly, decreases from run 3 to run 10,
  due to increased $\Rey$
    \citep{Linkmann2015PRL,Linkmann2017PRE, Bandyopadhyay2018PRX}.
  However, the `lengthscale'
  similarity constant, $\beta^{+}$, appears to be less sensitive to
  Reynolds number.}

{{Interestingly,}
K\'arm\'an--Howarth-like similarity decay has been
observed in 2DMHD \citep{Biskamp2001PoP}.  However, a comparison of
similarity solutions in 2DMHD
     (or 2.5D MHD)
    and strong-mean-field 3DMHD  is not entirely straightforward
    since 2DMHD also admits an inverse cascade of mean-squared
    magnetic potential, $A$.
      This requires that some magnetic energy is also inverse cascaded
      and is thus not available for direct cascade to the dissipative small
    scales.
    Exploring that parameter space is beyond the scope
    of the current paper.  In particular, the 2D runs would need to
    scan $E/A$ ($E$ is the energy) and the 3D ``comparison'' runs
    would require a scan of $B_0$, as well as varying the
    initial polarization (2D versus ``2.5D'').}

{Further, using $2.5\mathrm{D}$ fully kinetic
  Particle-In-Cell (PIC) simulations, it has been shown that weakly
  collisional plasmas support similarity decay
        \citep{Wu2013PRL, Parashar:ApJL2015}.}
It will be
interesting to extend and test the similarity decay phenomenologies
discussed here to three-dimensional kinetic simulations, shear driven
flows, compressible plasmas, etc.
{It is not clear why
  the quantity $\gamma = L_{\parallel} / L_{\perp}$ attains a constant
  value.  Other types of turbulent flow that develop anisotropy, due to
  rotation, stratification, convection, etc., may also admit a similar
  stabilization of the ratio of the parallel and perpendicular length
  scales.}
Another interesting direction in
which the similarity solution can be extended is quasi-static MHD
turbulence
        (see \cite{Verma2017RPP} for a review).
We plan to take up these endeavours in the future.

        \section*{Acknowledgments}
This research is supported by NASA under grant NNX17AB79G, and
Heliospheric Supporting Research grants 80NSSC18K1210, and
80NSSC18K1648, as well as the Parker Solar Probe
Project under Princeton subcontract SUB000016S.   The simulations were
performed using the \textit{Dante} and \textit{Casella} clusters at the
University of Delaware, USA.

\bibliographystyle{jfm}

\end{document}